\documentclass[showpacs,aps,nofootinbib,superscriptaddress, twocolumn, usenatbib,useAMS,10pt]{revtex4-1}
\input epsf
\usepackage{graphics}
\usepackage{amsmath}
\usepackage{amssymb}
\usepackage{bm}
\usepackage{braket}
\usepackage{booktabs}
\usepackage{subfigure}
\usepackage{subfloat}
\usepackage{float}
\usepackage[utf8]{inputenc}

\usepackage{hyperref}

\hypersetup{
    colorlinks=true,
    linkcolor=red,
    citecolor=blue,
}

\usepackage{color}
\usepackage{dcolumn}
\usepackage{hyphenat}

\def\be{\begin{equation}}
\def\ee{\end{equation}}
\def\ba{\begin{eqnarray}}
\def\ea{\end{eqnarray}}

\usepackage{graphicx}

\begin{document}

\title{MGCAMB with massive neutrinos and dynamical dark energy}

\author{Alex Zucca}  \affiliation{Department of Physics, Simon Fraser University, Burnaby, BC, V5A 1S6, Canada}
\author{Levon Pogosian} \affiliation{Department of Physics, Simon Fraser University, Burnaby, BC, V5A 1S6, Canada} \affiliation{Institute of Cosmology and Gravitation, University of Portsmouth, Portsmouth, PO1 3FX, UK}
\author{Alessandra Silvestri} \affiliation{Institute Lorentz, Leiden University, PO Box 9506, Leiden 2300 RA, The Netherlands}
\author{Gong-Bo Zhao} 
\affiliation{National Astronomical Observatories, Chinese Academy of Sciences, Beijing, 100101, P.~R.~China} 
\affiliation{School of Astronomy and Space Science, University of Chinese Academy of Sciences, Beijing, 100049, P. R. China} 
\affiliation{Institute of Cosmology and Gravitation, University of Portsmouth, Portsmouth, PO1 3FX, UK}

\begin{abstract}
We present a major upgrade of MGCAMB, a patch for the Einstein-Boltzmann solver CAMB used for phenomenological tests of general relativity against cosmological datasets. This new version is compatible with the latest CosmoMC code and includes a consistent implementation of massive neutrinos and dynamical dark energy. The code has been restructured to make it easier to modify to fit the custom needs of specific users. To illustrate the capabilities of the code, we present joint constraints on the modified growth, massive neutrinos and the dark energy equation of state from the latest cosmological observations, including the recent galaxy counts and weak lensing measurements from the Dark Energy Survey, and find a good consistency with the $\Lambda$CDM model.
\end{abstract}

\maketitle

{\hypersetup{hidelinks}
\tableofcontents
}

\section{Introduction}

Since the discovery of cosmic acceleration two decades ago \citep{Perlmutter:1998np, Riess:1998cb}, explaining it has been one of the primary goals of cosmology. The broadly accepted working model, in which the accelerated expansion is driven by the cosmological constant $\Lambda$ and most of the remaining density is in cold dark matter (CDM), fits remarkably well a plethora of observations, such as the cosmic microwave background (CMB) anisotropies \citep{WMAP, Ade:2015xua}, baryon acoustic oscillations (BAO) \citep{Percival:2002gq,6df,Alam:2016hwk}, type Ia supernovae \citep{Conley:2011ku, Suzuki:2011hu}, galaxy clustering \citep{SDSS} and galaxy lensing \citep{Heymans:2012gg, Hildebrandt:2016iqg}. However, the nature of CDM remains unknown, and the observed value of $\Lambda$ requires technically unnatural fine-tuning to reconcile it with the large vacuum energy density expected from quantum field theory \citep{Burgess:2017ytm}. These questions, along with new opportunities for testing gravity on cosmological scales afforded by the current and upcoming surveys \cite{Abbott:2005bi, LSST, EUCLID}, have motivated extensive studies of extensions of General Relativity (GR) \cite{Silvestri:2009hh,Clifton:2011jh,Ishak:2018his}.

Gravitational potentials evolve differently in alternative gravity theories compared to $\Lambda$CDM, leading to different predictions for the growth of cosmic structures. At linear order in cosmological perturbations, one can search for modified growth patterns phenomenologically, by introducing two functions that parameterize the altered relations between the Newtonian potential and the curvature perturbation, and between the matter density contrast and the Newtonian potential \cite{Amendola:2007rr,Zhang:2007nk,Hu:2007pj,Bertschinger:2008zb,2010PhRvD..81h3534B,Pogosian:2010tj}. Modified Growth with CAMB (MGCAMB) \citep{Zhao:2008bn, Hojjati:2011ix} is a patch for the popular Einstein-Boltzmann solver CAMB \cite{Lewis:1999bs}, enabling the calculation of cosmological observables for a given form of  such phenomenological functions in manner suitable for constraining them with CosmoMC \cite{Lewis:2002ah, Lewis:2013hha} or other similar Monte-Carlo Markov Chain algorithms.

The original version of MGCAMB \citep{Zhao:2008bn} and its update in \cite{Hojjati:2011ix} were largely based on the assumption that modifications of gravity appear well after the radiation-matter equality, and that the role played by the anisotropic stress in relativistic particle species is negligible. This limited the accuracy of modelling the effects of massive neutrinos. On the other hand, the neutrino mass can no longer be neglected in cosmological predictions, as the upcoming surveys are expected to probe masses close to the measured difference of $0.05$ eV between the masses of different neutrino flavours. Massive neutrinos contribute to the expansion rate as matter, but stream out of smaller gravitational potentials, suppressing the growth on small scales. This effect can be partially degenerate with those of modifications of gravity and, therefore, must be accounted for.

Another limitation of the prior versions of MGCAMB was the assumption of constant dark energy density. Generally, modified gravity (MG) theories predict modifications of the expansion history along with the modified growth of structures. The ability to study the covariance of the two can be important for ruling out broad classes of alternative gravity theories \cite{Espejo:2018hxa}.

This paper presents a major update of MGCAMB\footnote{This new patch is available at \url{https://github.com/sfu-cosmo/MGCAMB}} allowing for dynamical dark energy and accurate modelling of massive neutrinos. This version is compatible with the latest CosmoMC and is restructured to make it easier to customize to work with different parameterizations of phenomenological functions. The modifications of gravity can now be introduced at arbitrarily high redshifts, as long as the phenomenological functions continuously approach their $\Lambda$CDM values in the past. Users wishing to study models that do not approach the GR limit in the past will need to introduce the corresponding changes in the initial conditions.

To demonstrate the capabilities of this new version, we present joint constraints on the modified growth, massive neutrinos and the dark energy equation of state from the latest cosmological observations, including the Planck 2015 CMB data \cite{Aghanim:2015xee}, the Joint Light Curve analysis (JLA) supernovae \citep{Betoule:2014frx}, the BAO measurements from 6dF \citep{2011MNRAS.416.3017B} and SDSS DR7 \citep{Ross:2014qpa}, the measurements of the Hubble parameter, the angular-diameter distance and the redshift space distortions from BOSS DR12 \citep{Alam:2016hwk}, and the recent galaxy clustering and weak lensing measurements from the Dark Energy Survey (DES) \citep{Abbott:2017wau}.

The paper is organized as follows. In Section~\ref{sec:framework}, we briefly review the framework for phenomenological tests of gravity on cosmological scales. In Section~\ref{sec:patch}, we detail the implementation of the MGCAMB patch and the accuracy tests. Then, in Section~\ref{sec:analysis}, we demonstrate the use of MGCAMB by deriving joint constraints on modified gravity, massive neutrinos and the dark energy equation of state from the latest cosmological datasets. We conclude with a discussion in Section~\ref{sec:conclusion}.

\section{Modified Growth framework}
\label{sec:framework}

When describing deviations from GR on large scales, we assume that the universe is well described by a FRW metric with small perturbations. We adopt the conformal Newtonian gauge and write the line element as
\begin{equation}
ds^2 = a(\tau)^2 \left[ - (1 + 2 \Psi) d \tau^2 + (1-2 \Phi) dx^2 \right],
\end{equation}
where $\Psi$ and $\Phi$ are the scalar gravitational potential and the curvature perturbation, respectively, and both depend on the conformal time $\tau$ and the comoving coordinate ${\mathbf x}$.
We consider perturbations of the total energy-momentum tensor $T^{\mu}_{\nu}$ denoted as
\begin{align}
T^0_0 + \delta T^0_0 & = -\rho (1 + \delta), \\
T^0_i + \delta T^0_i & = -(\rho + P)v_i , \\
T^i_j + \delta T^i_j & = (P + \delta P ) \delta^i_j + \pi^i_j,
\end{align}
where $\delta$ is the density contrast, $v$ is the velocity field, $\delta P$ the pressure perturbation and $\pi^i_j$ denotes the traceless anisotropic stress tensor. The energy momentum tensor components evolve according to the conservation equation, $T^{\mu}_{\nu ; \mu} = 0$; working in Fourier space, the perturbations obey \cite{Bertschinger:2006aw}
\begin{align}
\label{eq:delta_dot}
\dot{\delta} & = - (1+w) \left( \theta - 3 \dot{\Phi} \right) - 3 \mathcal{H} \left( \frac{\delta P}{\delta \rho} - w \right) \delta, \\
\label{eq:theta_dot}
\dot{\theta} & = - \mathcal{H}(1 - 3 w) \theta - \frac{\dot{w}}{1+w} \theta  + \frac{\delta P / \delta \rho}{1+w} k^2 \delta - k^2 \sigma + k^2 \Psi\,,
\end{align}
where $\theta$ is the divergence of the velocity field. These equations hold for the combination of all cosmological fluids or for any decoupled subset of fluids such as CDM, the photon-baryon fluid and massive neutrinos after decoupling. 

To close the system of equations for cosmological perturbations, one needs two additional equations relating the perturbed energy momentum tensor and the metric potentials $\Phi$ and $\Psi$. In $\Lambda$CDM, one can combine the ${}^0_0$ and the divergence of the ${}^0_i$ Einstein equations  to obtain 
\begin{equation}
k^2 \Phi = - 4 \pi G a^2 \rho \Delta,
\label{eq:Phi-Delta}
\end{equation}
where $\Delta$ is the gauge-invariant density contrast,
\begin{equation}
\rho \Delta \equiv \rho \delta + \frac{3 \mathcal{H}}{k^2} (\rho + P)\theta.
\end{equation}
One can also use the traceless part of the ${}^i_j$ Einstein equation, given by
\begin{equation}
k^2 (\Phi - \Psi) = 12 \pi G a^2 (\rho + P) \sigma,
\label{eq:grav_slip_GR}
\end{equation}
and use this to write an equation relating $\Psi$ and $\Delta$:
\begin{equation}
k^2 \Psi = - 4 \pi G a^2 \left[ \rho \Delta + 3 (\rho+P) \sigma \right].
\label{eq:Poisson_GR}
\end{equation}
Eqs.~(\ref{eq:grav_slip_GR}) and (\ref{eq:Poisson_GR}), combined with the conservation equations \eqref{eq:delta_dot} and \eqref{eq:theta_dot}, can be used in Einstein-Boltzmann solvers, such as CAMB and CLASS, to compute the cosmological observables. 

Einstein's equation would be modified in alternative gravity theories. For example, in scalar-tensor theories of gravity, the scalar field alters the relation between $\Phi$ and $\Psi$, and mediates an additional force between massive particles, which enhances the growth of structure. This can be phenomenologically modelled by introducing two functions of time and scale, $\mu(a,k)$ and $\gamma(a,k)$, that encode possible modification of (\ref{eq:grav_slip_GR}) and (\ref{eq:Poisson_GR}), defined via
\ba
\label{eq:mu_equation}
k^2 \Psi = - 4 \pi G  \mu(a, k) a^2 \left[ \rho \Delta + 3 (\rho+P) \sigma \right], \\
\label{eq:gamma_equation}
k^2 [\Phi - \gamma(a,k) \Psi] = 12 \pi G \mu(a,k) a^2 (\rho + P) \sigma.
\ea
Given the form of $\mu(a,k)$ and $\gamma(a,k)$,  Eqs~\eqref{eq:mu_equation} and \eqref{eq:gamma_equation}, along with  \eqref{eq:delta_dot} and \eqref{eq:theta_dot}, can be used to evolve the system of equations and compute the cosmological observables of interest. Note that $\mu(a,k)$ is introduced as a modification of Eq.~(\ref{eq:Poisson_GR}) that relates $\Psi$ and $\Delta$, instead of \eqref{eq:Phi-Delta}, because matter perturbations respond to the gradients of the Newtonian potential $\Psi$. This makes $\mu$ a parameter that directly controls the strength of the gravitational interaction.

The gravitational slip function $\gamma$ is not directly related to cosmological observables \citep{Amendola:2007rr, Amendola:2012ky} and, therefore, is generally difficult to constrain without fixing $\mu$ or making additional assumptions. Instead, it is often more informative to work with $\mu$ paired with function $\Sigma$ that modifies the relation between the lensing potential $(\Phi + \Psi)$ and $\Delta$ via
\begin{equation}
k^2 (\Phi + \Psi) = - 4 \pi G \Sigma(a,k) a^2  \left[ 2\rho \Delta + 3(\rho+P) \sigma \right].
\label{eq:Sigma}
\end{equation}
In the limit of negligible anisotropic stress, $\Sigma$ is simply related to $\mu$ and $\gamma$ through $\Sigma = \mu (1+\gamma)/2$. One can break the degeneracy between $\mu$ and $\Sigma$ and constrain both of them independently by combining data from clustering surveys with measurements of weak lensing \cite{WLRSD,Simpson13}. While it might be obvious to many readers, we would still like to note that constraints on $\mu$ depend on whether one marginalizes over $\gamma$ or $\Sigma$.

The above-mentioned framework is implemented in MGCAMB, which can be used for two purposes. One can adopt functional forms of $\mu$ and $\gamma$, or $\mu$ and $\Sigma$, and fit the function parameters to data to search for any departure from the $\Lambda$CDM values of $\mu=\gamma=\Sigma=1$. One can go even further and reconstruct the functions from the data using the correlated prior approach \cite{CP1,CP2,CPDE,CPX}, or the Gaussian Process Reduction \cite{GP1,GP2}. Another way to use MGCAMB is to study predictions of specific theories for certain cosmological observables. This application is limited by the fact that deriving the analytical forms for $\mu$, $\gamma$ and $\Sigma$ in a given theory requires adopting the quasi-static approximation (QSA) \cite{Silvestri:2013ne}. The validity of the QSA depends on the observable and the strength of modifications introduced by the theory on near-horizon scales. As a rule, QSA tends to work in most viable models \cite{Silvestri:2013ne,Sawicki:2015zya} and thus MGCAMB can be a good starting point for looking at the characteristic observational signatures of a theory.

\section{The new MGCAMB patch}
\label{sec:patch}

The set of equations used in the new MGCAMB patch is based on and has a large overlap with the previous versions \citep{Zhao:2008bn, Hojjati:2011ix}. However, there are several important differences and, for the sake of completeness, we present the entire formalism.

In CAMB, cosmological perturbations are evolved in synchronous gauge, with the line element given by \citep{Bertschinger:2006aw}
\begin{equation}
ds^2 = a(\tau)^2 \left[ - d \tau^2 + (\delta_{ij}+h_{ij}({\mathbf x},\tau)) dx^i dx^j \right],
\end{equation}
where 
\be
h_{ij}=\int d^3k e^{i {\mathbf k} \cdot {\mathbf x}} [ \hat{\mathbf k}_i  \hat{\mathbf k}_j h({\mathbf k},\tau) + (\hat{\mathbf k}_i \hat{\mathbf k}_j - \delta_{ij}/3) 6\eta({\mathbf k},\tau)].
\ee
The Newtonian gauge potentials $\Phi$ and $\Psi$ are related to the synchronous gauge potentials $\eta$ and $h$ through
\begin{align}
\label{eq:phi_to_synch}
\Phi & =\eta - \mathcal{H}\alpha, \\
\label{eq:psi_to_synch}
\Psi & =\dot{\alpha} + \mathcal{H}\alpha,
\end{align}
where $\alpha = (\dot{h}+6 \dot{\eta})/2k^2$. The synchronous gauge perturbations of the energy-momentum tensor evolve according to
\begin{align}
\label{eq:deta_dot_synch}
\dot{\delta} & = - (1+w) \left(\theta + \frac{\dot{h}}{2}\right) - 3 \mathcal{H}\left( \frac{\delta P }{\delta \rho} - w \right) \delta, \\
\label{eq:theta_dot_synch}
\dot{\theta} & = -\mathcal{H}(1 - 3 w) \theta - \frac{\dot{w}}{1+w} \theta + \frac{\delta P / \delta \rho}{1+w} k^2 \delta - k^2 \sigma.
\end{align}
In order to evolve the perturbations, one needs to compute the quantity $\dot{h}$, or the quantity $\mathcal{Z}$ defined in CAMB as $\mathcal{Z} \equiv \dot{h}/2 k$. In CAMB, this is done using the ${}^0_0$ Einstein equation in synchronous gauge
\begin{equation}
k^2 \eta + k \mathcal{H} \mathcal{Z} = - 4 \pi G a^2 \rho \delta.
\end{equation}
In MGCAMB, the Einstein equations are modified and, hence, one needs an alternative way to compute $\mathcal{Z}$. From the definition of $\alpha$, we have
\be
\mathcal{Z} = k \alpha - {3\dot{\eta} \over k}.
\label{eq:zeta}
\ee
To find $\alpha$ and $\dot{\eta}$, we start by substituting Eqs.~\eqref{eq:phi_to_synch} and \eqref{eq:psi_to_synch} into the modified Einstein equations \eqref{eq:mu_equation} and \eqref{eq:gamma_equation}, and combine the resulting equations to write the following expression for $\alpha$:
\begin{equation}
\label{eq:Alpha}
\alpha = \left\{ \eta + \frac{\mu a^2}{2 k^2} \left[ \gamma\rho\Delta + 3(\gamma-1)(\rho+P) \sigma \right] \right\} \frac{1}{\mathcal{H}},
\end{equation}
where we now include the factor $8 \pi G$ in the definition of density and pressure, e.g. $ 8 \pi G \rho \to \rho$.
To derive an equation for $\dot{\eta}$, we first rearrange the equation above to solve for $\eta$, obtaining
\begin{equation}
\label{eq:eta}
\begin{split}
\eta & = \mathcal{H} \alpha - \frac{\mu a^2}{2  k^2} \left\{ \gamma \rho \Delta + 3 (\gamma-1) \rho (1+w) \sigma \right\} \\
& = \mathcal{H} \alpha - \frac{\mu a^2}{2  k^2} \Gamma,
\end{split}
\end{equation}
where we defined 
\begin{equation}
\Gamma = \gamma \rho \Delta + 3 (\gamma-1)\rho(1+w) \sigma.
\end{equation}
Next, we would like to differentiate Eq.~\eqref{eq:eta} with respect to $\tau$. To compute $(\rho \Delta)^{\cdot}$, we combine the conservation equations \eqref{eq:deta_dot_synch} and \eqref{eq:theta_dot_synch} to obtain
\begin{equation}
\label{eqn:rhoDeltadot}
\begin{split}
(\rho \Delta)^{\cdot} & = - 3 \mathcal{H} \rho \Delta - (1+w) \rho \theta \left[ 1+ \frac{3}{k^2} (\mathcal{H}^2  - \dot{\mathcal{H}}) \right] \\ 
& - 3 \mathcal{H}\rho (1+w) \sigma - (1+w)\rho k \mathcal{Z}.
\end{split}
\end{equation}
Finally, taking the derivative of Eq.~\eqref{eq:eta} and using Eqs.~\eqref{eq:zeta} and \eqref{eqn:rhoDeltadot}, leaves us with the equation for $\dot{\eta}$:
\begin{equation}
\begin{split}
\dot{\eta}  = & \frac{1}{2} \frac{a^2}{3\rho a^2 \mu \gamma (1+w)/2 + k^2} \\
& \times \biggl\{ \rho (1+w) \mu \gamma \theta \left[ 1 + 3 \frac{\mathcal{H}^2 - \dot{\mathcal{H}}^2}{k^2}\right] \\
&  + \rho \Delta \left[ \mathcal{H} \mu (\gamma - 1) - \dot{\mu} \gamma - \mu \dot{\gamma} \right] \\
& + 3 \mu (1-\gamma) \rho (1+w)\dot{\sigma} \\ 
& + k^2\alpha \left[ \rho \mu \gamma (1+w)  - 2 \left( \frac{\mathcal{H}^2 - \dot{\mathcal{H}}}{a^2} \right)\right]  \\
& + 3\mathcal{H}\mu  (\gamma - 1) (1+w) \rho \sigma (3w+2)\\
& - 3 (1+w) \rho \sigma  \left[\dot{\mu} (\gamma - 1) - \dot{\gamma} \mu  + \mu (1-\gamma) \frac{\dot{w}}{1+w} \right] \biggr\},
\end{split}
\end{equation}
where we replaced $\dot{\alpha}$ with
\begin{equation}
\dot{\alpha} = \Phi + \Psi - \eta,
\end{equation}
and used the modified Einstein equations to express $\Phi$ and $\Psi$ in terms of the energy-momentum perturbations.
 
Following the notation in CAMB, we introduce fluxes $q$ and the anisotropic stress $\Pi$ related to the velocity divergence $\theta$ and the anisotropic stress $\sigma$ through
\begin{equation}
(1+w)\theta = k q, \quad \frac{3}{2}(1+w)\sigma = \Pi,
\end{equation}
to rewrite the equation for $\dot{\eta}$ as
\begin{equation}
\begin{split}
\dot{\eta} & = \frac{1}{2} \frac{a^2}{3 \rho a^2 \mu \gamma (1+w)/2 + k^2} \biggl\{ \rho  \mu \gamma k q \left[ 1 + 3 \frac{\mathcal{H}^2 - \dot{\mathcal{H}}}{k^2} \right]  \\
& + \rho \Delta \left[ \mathcal{H} \mu (\gamma - 1) - \dot{\mu} \gamma - \mu \dot{\gamma} \right] \\
& + 2 \mu (1-\gamma) \rho \dot{\Pi} \\
& + k^2\alpha \left[  \rho \mu \gamma (1+w) - 2 \left( \frac{\mathcal{H}^2 - \dot{\mathcal{H}}}{a^2} \right)  \right] \\
& + 2 \rho \Pi \left[ \mathcal{H} (\gamma - 1)(3w+2)\mu - \dot{\mu} (\gamma - 1) - \dot{\gamma} \mu   \right] \biggr\}.
\end{split}
\label{eq:etadot}
\end{equation}
The background and perturbation variables of energy-momentum appearing in the above equations are sums over the uncoupled fluid components, e.g. $\rho = \rho_{\rm b+\gamma} + \rho_{\rm CDM}  + \rho_{\nu}$, {\it etc}.

There are two notable differences between Eq.~\eqref{eq:etadot} and its counterpart in the previous MGCAMB patch. First, the factor $(3w+2)$ in the last line corrects a typo present in the previous version\footnote{This factor is $(3w)$ in the last term of Eq.~(36) of \citep{Hojjati:2011ix}.}. As this correction is proportional to $\Pi$, it has a negligible effect at late times when the anisotropic stress is small. More importantly, the pre-factors of $\alpha$ and $q$ are now generalized to allow for an arbitrary expansion history. The expression for $\dot{\eta}$ in the previous version of MGCAMB assumed dark energy with the equation of state $w_{\rm DE}=-1$.
 
As in previous versions, the present MGCAMB patch assumes that GR is recovered deep in the radiation era. The code starts with the same initial conditions as CAMB and evolves the original CAMB system of equations up to a certain value of the scale factor, $a_{\rm trans}$, set by the parameter \verb!GRtrans!. After that, the code evolves the alternative equations described above. Unlike the previous version, the present patch has no restriction on how early the switch can happen, as long as it happens after the time at which the initial conditions are set for the smallest values of $k$ and the phenomenological functions are such that the GR limit is approached continuously.

Computing $\mathcal{Z}$ requires knowing the quantities $\delta$, $q$, $\Pi$ and $\dot{\Pi}$, which can be a challenging problem depending on the epoch at which $\dot{\eta}$ is evaluated. For example, at late times, CAMB stops evolving the full set of Boltzmann equations for photons and neutrinos and uses the radiation streaming approximation (RSA) instead \citep{2011JCAP...07..034B}. In the RSA, $\delta_{\gamma}$ and $\delta_{\nu}$ are computed by using approximated versions of $\mathcal{Z}$ and $\dot{\mathcal{Z}}$ that do not include radiation. However, the current CAMB implementation of RSA uses the ${}^i_i$ Einstein equation. Since we do not have all the modified Einstein equations, we opt to use the RSA implementation from an older version of CAMB, which did not depend on the ${}^i_i$ Einstein equation. For small values of \verb!GRtrans!, in order to preserve the accuracy we had to increase the time at which the RSA is switched on, which slows down MGCAMB with respect to the default CAMB by a factor of two.

Before last scattering, CAMB uses a second order tight coupling expansion in which the computation of $q_{\gamma}$ and $q_b \equiv v_b$ requires the knowledge of $\mathcal{Z}$ and $\sigma^* \equiv  k \alpha$. We resolve this by using the values of these quantities computed at the previous time-step.

In addition to the ($\mu,\gamma$) parameterization, MGCAMB offers options to work with ($\mu,\Sigma$) introduced in Sec.~\ref{sec:framework} and the ($Q,R$) functions of \citep{2010PhRvD..81h3534B}. More details on their implementation are given in Appendix~\ref{sec:other_parameterizations}. 

\subsection{Massive Neutrinos}

\begin{figure*}[tbh]
\centering
\includegraphics[width=.9\textwidth]{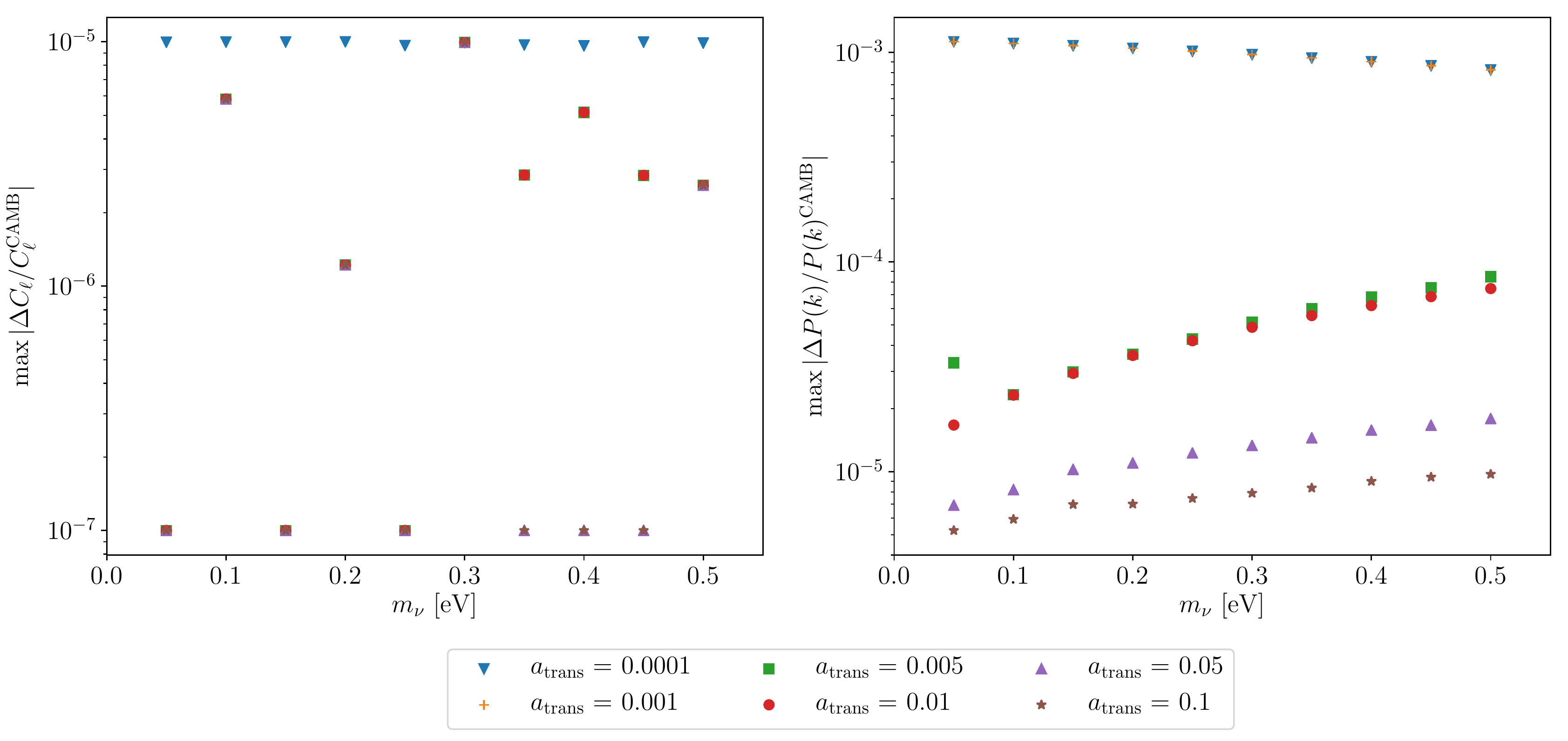}
\caption{Maximum relative difference in $C_\ell^{\rm TT}$ and $P(k)$ between the GR limit ($\mu=\gamma=1$) of MGCAMB and standard CAMB for several values of the sum of the neutrino masses and different values of the scale factor at which the modified equations are turned on.}
\label{fig:MGCAMB_test}
\end{figure*}

The default CAMB code calculates the quantities $\dot{\Pi}$ after computing $\mathcal{Z}$. In our case, $\dot{\Pi}$ is required to compute $\mathcal{Z}$. The computation of $\dot{\Pi}_{\gamma}$ and $\dot{\Pi}_{r}$ in the previous versions of MGCAMB ignored the contribution from massive neutrinos. In the current version, we compute $\dot{\Pi}_{\nu}$ by integrating the neutrinos equations before the computation of $\mathcal{Z}$. This is safe, since the equations for $\dot{\Pi}_{\nu}$ do not depend on $\mathcal{Z}$.

\subsection{The CMB source function and the weak lensing transfer function}
The CMB temperature angular power spectrum is given by \citep{Zaldarriaga:1996xe}
\begin{equation}
C_{\ell}^{\rm TT} = (4 \pi)^2 \int dk \, k^2 |\Delta_{\ell}^{\rm T}(k)|^2 P_{\cal R}(k),
\end{equation}
where $P_{\cal R}(k)$ is the primordial curvature perturbation power spectrum and the CMB temperature transfer function is given by
\begin{equation}
\Delta_{\ell}^{\rm T}(k) = \int_0^{\tau_0} d \tau S_{\rm T}(k, \tau) j_{\ell}(k \tau),
\end{equation}
where $ S_{\rm T}(k, \tau)$ is the source term and $j_{\ell}(x)$ are the spherical Bessel functions. In synchronous gauge, the source is
\begin{equation}
\begin{split}
S_{\rm T}(k,\tau) = & g \left( \Delta_{{\rm T} 0} + 2 \dot{\alpha} + \frac{\dot{v}_b}{k} + \frac{\Pi^{\rm pol}}{4} + \frac{3 \ddot{\Pi}^{\rm pol}}{4 k^2} \right) \\
& + e^{- \kappa}(\dot{\eta} + \ddot{\alpha}) + \dot{g} \left( \alpha + \frac{v_b}{k} + \frac{3 \dot{\Pi}^{\rm pol}}{2 k^2} \right) \\
& +  \frac{3 \ddot{g}\Pi^{\rm pol}}{4 k^2},
\end{split}
\end{equation}
where $\kappa$ is the optical depth, $g$ is the visibility function and $\Pi^{\rm pol}$ is the polarization term. In CAMB, the calculation of the ISW term, $(\dot{\eta} + \ddot{\alpha})$, assumes GR and has to be replaced in MGCAMB. This was already done in the previous versions, but the contribution of massive neutrinos was neglected. We introduce massive neutrinos properly in the current version, using the following prescription. The ISW term can be written as
\begin{equation}
\ddot{\alpha} + \dot{\eta} = \dot{\Phi} + \dot{\Psi},
\end{equation}
where terms on the right hand side can be computed separately. The first term is determined by taking a derivative of Eq.~\eqref{eq:phi_to_synch}, giving
\begin{equation}
\dot{\Phi} = \dot{\eta} - \mathcal{H} (\Psi - \mathcal{H} \alpha) - \dot{\mathcal{H}} \alpha.
\end{equation}
Then, $\dot{\Psi}$ can be obtained by differentiating the modified Poisson equation \eqref{eq:mu_equation},
\begin{equation}
\label{eqn:Psi_dot}
\begin{split}
\dot{\Psi} & = - \frac{1}{2 k^2} \dot{\mu} a^2 \left[ \rho \Delta + \rho \Pi \right] \\
& - \frac{1}{2 k^2} \mu \left[ (\rho a^2 \Delta)^{\cdot} + 2 (\rho a^2 \Pi)^{\cdot} \right],
\end{split}
\end{equation}
where $(\rho a^2 \Delta)^{\cdot}$ is determined from Eq.~\eqref{eqn:rhoDeltadot} and
\begin{equation}
(\rho a^2 \Pi)^{\cdot} = \rho a^2 \dot{\Pi} - \left[ 2 {\cal H} \rho a^2 \Pi + (3 P - \rho )a^2 \Pi \right],
\end{equation}
completing the set of required equations in a form suitable for implementation in CAMB.

The last modification concerns the implementation of the weak lensing transfer function, since CAMB assumes GR to calculate the Weyl potential $\Phi+\Psi$. Instead, in MGCAMB, we compute $\Psi$ and $\Phi$ from Eqs.~\eqref{eq:mu_equation} and \eqref{eq:gamma_equation}. The Weyl potential is used to compute the lensing correlation functions in the analysis presented in Section~\ref{sec:analysis}.

\subsection{The GR limit of MGCAMB}
We have checked the output of MGCAMB in the GR limit, when $\mu=\gamma=1$, for a wide range of neutrino masses and values of the scale factor at which modifications are switched on. Fig.~\ref{fig:MGCAMB_test} shows the maximum relative difference in $C_\ell^{\rm TT}$ and $P(k)$ between CAMB and MGCAMB for $0.05$ eV $\le \sum m_{\nu} \le 0.5$ eV and $0.0001\le a_{\rm trans}\le 0.1$. In all cases the deviations are below 0.1\%. In order to achieve this accuracy for $a_{\rm trans}<0.005$, we had to delay the time at which the RSA in MGCAMB is switched on by a factor of 20, which doubles the running time of the code. 

\section{Joint constraints on modified growth, massive neutrinos and the dark energy equation of state}
\label{sec:analysis}

The MGCAMB patch has been implemented in the Markov Chain Monte Carlo (MCMC) engine CosmoMC \cite{Lewis:2002ah, Lewis:2013hha}, and is called MGCosmoMC\footnote{Available at \url{https://github.com/sfu-cosmo/MGCosmoMC}}. To demonstrate its use, we derive joint constraints of massive neutrinos, modified growth, and the DE equation of state from the datasets included in the current version of CosmoMC. We consider three models: $\Lambda$CDM with the sum of  neutrino masses $\sum m_\nu$ as an additional parameter (hereafter referred to as Model 0), a model with $\mu$ and $\gamma$ varied along with $\sum m_\nu$ and $\Lambda$ playing the role of dark energy (Model 1), and a model with the DE equation of state varying in addition to $\sum m_\nu$, $\mu$ and $\gamma$ (Model 2). 

In Models 1 and 2, we adopt the ($\mu,\gamma$) parameterization used by the Planck collaboration \citep{Ade:2015rim,Aghanim:2018eyx}, {\it i.e.}
\begin{align}
\mu(a) & = 1 + E_{11} \Omega_{\rm DE} (a), \\
\gamma(a) & = 1 + E_{21} \Omega_{\rm DE}(a),
\end{align}
where $\Omega_{\rm DE}(a)=\rho_{\rm DE}/\rho_{\rm tot}$. We present the results in terms of the derived quantities $\mu_0 \equiv \mu (a = 1)$, $\gamma_0 \equiv \gamma (a = 1)$ and $\Sigma_0 \equiv \mu_0 (1+\gamma_0)/2$. In Model 2, we adopt the CPL parameterization \citep{Chevallier:2000qy, Linder:2002et} of the DE equation of state,
\begin{equation}
w_{\rm DE}(a) = w_0 + (1-a) w_a.
\end{equation}
In all cases, we also varied the six ``vanilla'' $\Lambda$CDM parameters, running four parallel chains until the Gelman/Rubin convergence statistics reached $R<0.01$.

\subsection{The datasets}

Our analysis made use of the Planck CMB temperature and polarization anisotropy spectra in combination with other datasets.  Although the 2018 results were recently released \cite{Aghanim:2018eyx}, the latest Planck likelihood code was not available at the time of writing, so we used the 2015 version \cite{Aghanim:2015xee}. Specifically, we used the Planck 2015 TT, TE and EE likelihood for multipoles in the range $30 \le \ell \le 2508$ along with the lowTEB polarization likelihood for multipoles in the range $2 \le \ell \le 29$.  We also used the Planck CMB lensing measurements from the minimum variance combination of temperature and polarization with the conservative cut\footnote{We use the files \url{smicadx12_Dec5_ftl_mv2_ndclpp_p_teb_consext8}} of $40 \le \ell < 400$. 
 
We have combined CMB data with the Type Ia supernovae luminosity distance measurements from JLA \citep{Betoule:2014frx}, the BAO measurements from the 6dF galaxy survey \citep{2011MNRAS.416.3017B} (at $z = 0.106$), the SDSS DR7 Main Galaxy Sample (MGS) \citep{Ross:2014qpa} (at $z = 0.15$), the measurements of the Hubble parameter $H(z_i)$, the angular-diameter distance $d_A(z_i)$ and the redshift space distortion measurements of $f(z_i)\sigma_8(z_i)$ at $z_{i}= \{0.38, 0.51, 0.61\}$ provided by the BOSS DR12 \citep{Alam:2016hwk}. As usually done in the literature, we assumed that the 6DF and MGS measurements are independent from BOSS DR12.

We have also used the recent galaxy clustering and weak lensing measurements from the DES Year 1 results \citep{Abbott:2017wau}. This dataset consists of the measurements of the angular two-point correlation functions of galaxy clustering, cosmic shear and galaxy-galaxy lensing in a set of 20 logarithmic bins of angular separation in the range $2.5^{\prime} - 250^{\prime}$. Since the MG formalism has no nonlinear prescription for structure formation, the angular separations probing the nonlinear scales were properly removed. To do so, we adopted the same method as in \citep{Abbott:2018xao} and used the ``standard'' data DES cutoff as described in Appendix~\ref{sec:linear_DES_data}. Moreover, we modified the DES likelihood code in order to compute the theoretical predictions of the cosmic shear and the galaxy-galaxy lensing correlation functions using the Weyl potential $k^2(\Phi + \Psi)/2$ instead of using the GR approximation $k^2(\Phi + \Psi)/2 =k^2 \Phi = -(3/2)\Omega_m H_0^2 a^{-1} \delta_m $, as described in Appendix~\ref{sec:weyl_like}. Finally, the covariance between the DES data and the 6DF, MGS and BOSS measurements is ignored, as the observations are carried on different sky patches \citep{Abbott:2018xao}.

\subsection{The GR limit consistency check}
\label{sec:GRconsistency}

\begin{table*}[tbh]
\centering
\begin{tabular}{l|c|c|c}
\hline
\hline
Parameter & CosmoMC  & MGCosmoMC ($a_{\rm trans} = 0.01$) & MGCosmoMC ($a_{\rm trans} = 0.001$) \\
\hline
$\omega_b $                      & $0.02237 \pm 0.00014$ & $0.02237 \pm 0.00014$  &  $0.02237 \pm 0.00014$\\
$\omega_c $                      & $0.1178 \pm 0.0011$   & $0.1178 \pm 0.0011$    &  $0.1178 \pm 0.0011$  \\
$100 \theta_{\rm MC}$            & $1.04095 \pm 0.00030$ & $1.0410 \pm 0.0003$&  $0.104095 \pm 0.00030$\\
$\tau$                           & $0.076 \pm 0.015$     & $0.075 \pm 0.015$  &  $0.075 \pm 0.015$ \\
$\sum m_{\nu}$   (95 \% CL)    & $< 0.206$ eV          & $< 0.198$ eV          &  $< 0.212$ eV \\
$n_s$                            & $0.9684  \pm 0.0042$  & $0.9684 \pm 0.0042$   &  $0.9684  \pm 0.0042$\\
$\ln \left( 10^{10} A_s \right)$ & $3.080 \pm 0.028$     &   $3.079 \pm 0.027$    &  $3.080 \pm 0.029$\\
\hline
Best fit: $ -\log ({\rm Like})$  & $7023.371$            &       $7023.607$       &  $7023.964$\\
\hline
\hline
\end{tabular}
\caption{The 68 \% CL uncertainties and best fit values of parameters obtained using the original CosmoMC, compared to the results from the GR limit of MGCosmoMC for two different values of $a_{\rm trans}$ which sets the scale factor beyond which the modified set of equations is evolved. The bound on the net mass of neutrinos is at the 95\% CL.}
\label{tab:gr_test_results}
\end{table*}

To assess the impact of the small systematic errors introduced by the approximations used in MGCAMB, we performed a consistency check of the GR limit of MGCosmoMC by comparing the results of three MCMC runs: 1) using the original CosmoMC code, 2) using MGCosmoMC with $\mu = \gamma = 1$ and $a_{\rm trans} = 0.01$ and 3) using MGCosmoMC with $\mu = \gamma = 1$ and $a_{\rm trans} = 0.001$. In all runs, we varied the six vanilla $\Lambda$CDM parameters and the mass of neutrinos. The results are summarized in Table~\ref{tab:gr_test_results}. We can see that the best fit values and the confidence intervals for cosmological parameters are practically the same in all cases and, hence, the results are consistent.

\subsection{Results}

\begin{table*}[tbh]
\centering
\begin{tabular}{l| c | c | c }
\hline
\hline
Parameter & Model 0 & Model 1 & Model 2 \\
\hline
$\omega_b $                     & $0.02237 \pm 0.00014$ & $0.02239 \pm 0.00014$ & $ 0.02231 \pm 0.00016 $ \\
$\omega_c $                     & $0.1178 \pm 0.0011$   & $0.1175 \pm 0.0011$   & $ 0.1183 \pm 0.0013 $ \\
$100 \, \theta_{\rm MC}$        & $1.0409 \pm  0.0003$  & $1.0410  \pm 0.0003$  & $ 1.0408 \pm 0.0003$\\
$\tau$                          & $0.075 \pm 0.015$     & $0.067 \pm 0.017$     & $ 0.072 \pm 0.018 $\\
$n_s$                           & $0.969 \pm 0.004$     & $0.969 \pm 0.004$     & $ 0.967 \pm 0.005$ \\
$\ln \left(10^{10} A_s \right)$ & $ 3.08 \pm 0.03$      & $3.06  \pm 0.03$      & $ 3.07 \pm 0.03$\\
\hline
$\sum m_{\nu}$ (95 \% CL)       & $< 0.21$ eV           & $<0.24$ eV            & $<0.49$ eV\\
$\mu_0 - 1$                     & $0$                   & $-0.09 \pm 0.30$      & $-0.07 \pm 0.29$\\
$\gamma_0 - 1$                  & $0$                   & $0.46 \pm 0.79$       & $ 0.43 \pm 0.77 $\\ 
$\Sigma_0 - 1$                  & $0$                   & $0.01 \pm 0.06$       & $ 0.02 \pm  0.07$\\
$w_0$                           & $-1$                  & $-1$                  & $ -0.84 \pm 0.13$\\
$w_a$                           & $0$                   & $0$                   & $ -0.48 \pm 0.36$\\
\hline
$\chi^2$                        & $  7023.37$           & $7023.04$             & $7024.44$ \\
$\Delta \chi^2$                 & -                     & $-0.33$               & $+1.07$\\
\hline
\hline
\end{tabular}
\caption{\label{tab:MG_mnu_DE_results} The 68 \% CL uncertainties and best fit values of parameters constrained using MGCosmoMC. The bound on the net mass of neutrinos is at the 95\% CL. Model 0 corresponds to $\Lambda$CDM with massive neutrinos. Model 1, in addition, includes modified growth on the $\Lambda$CDM background, while Model 2 adds a varying DE equation of state using the CPL parameterization.}
\end{table*}

\begin{figure}[tbh]
\centering
\includegraphics[width=.5\textwidth]{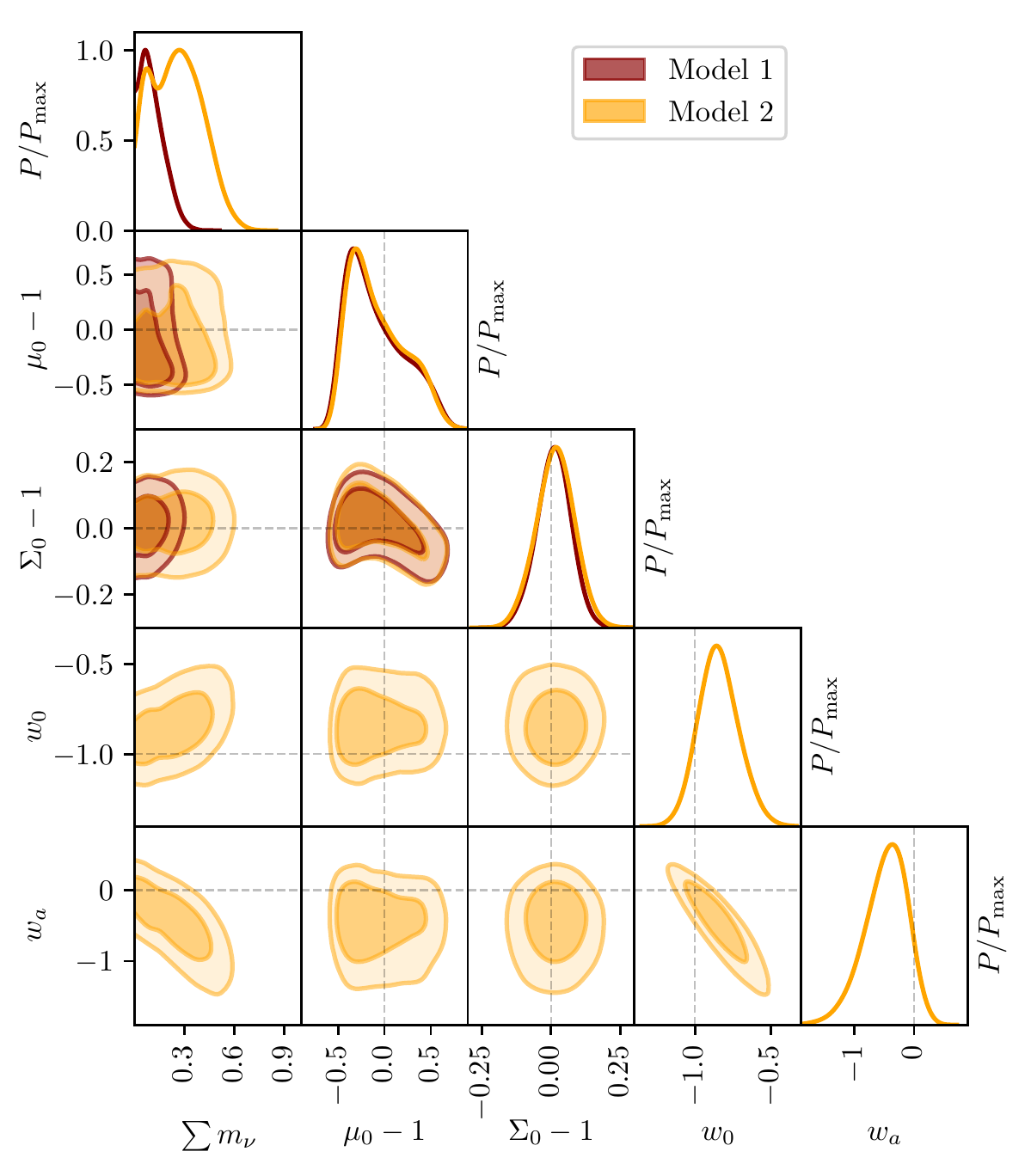}
\caption{\label{fig:triangular_plot} The marginalized joint posterior distribution of the Model 1 and Model 2 parameters. The plots along the diagonal show the marginalized posterior distribution of each parameter. The grey dashed lines indicate the $\Lambda$CDM limit values of the additional parameters. The darker and lighter shades correspond to the 68\% and the 95\% CL, respectively.}
\end{figure}

The joint constraints derived on massive neutrinos, modified growth and the DE equation of state are summarized in Table~\ref{tab:MG_mnu_DE_results} for Models 0, 1 and 2 defined at the beginning of this Section. Also, Fig.~\ref{fig:triangular_plot} shows the marginalized distributions of the relevant parameters, with their $\Lambda$CDM limits shown with dashed grey lines. We find the 95\% CL bound on massive neutrinos to be
\begin{align*}
\sum m_{\nu} & < 0.21 \, {\rm eV}, \quad {\rm Model} \, 0, \\
\sum m_{\nu} & < 0.24 \, {\rm eV}, \quad {\rm Model} \, 1, \\
\sum m_{\nu} & < 0.49 \, {\rm eV}, \quad {\rm Model} \, 2.
\end{align*}
Our bound on the neutrino mass for Model 0 is comparable to the DES Year 1 result of $0.29$ eV at 95\% CL \citep{Abbott:2017wau}. The use of the CMB polarization data at high-$\ell$ in our analysis is the reason for the stronger constraint.

In Model 1, the effective (cosmological) Newton's constant can vary at late times. Such variation can happen, for example, in scalar-tensor theories of gravity, where it would generally be scale-dependent. In such theories, the extra Yukawa force mediated by the scalar gravitational degree of freedom enhances the structure formation at scales below the Compton wavelength of the scalar field, which could negate the free streaming suppression of structure formation due to the non-zero neutrino mass. Since in our analysis we considered a scale-independent parameterization of $\mu,\gamma$, this degeneracy between $\sum m_{\nu}$ and $\mu_0-1$ is not present and the constraint on $\sum m_{\nu}$ is comparable to the one in Model 0.

In Model 2, $\mu$ is also scale-independent, however the DE density is time-dependent and the degeneracy between the dynamics of DE and the neutrino mass weakens the constraints on $\sum m_{\nu}$. The degeneracy between the neutrino mass and the CPL parameters $w_0$ and $w_a$ is evident from Fig.~\ref{fig:triangular_plot}. The 95\% C.L. bounds on the modified growth parameters are consistent with the $\Lambda$CDM limit and with the results obtained by DES \citep{Abbott:2018xao}. Note that, as expected, the bounds on $\gamma_0$ are generally weaker than those on $\mu_0$ and $\Sigma_0$, because there is no observable that can cleanly separate its effect from the latter two. Our constraints on the variation in the DE equation of state also indicate a good agreement with the $\Lambda$CDM model. 

The analysis presented in this Section illustrates how the new MGCAMB patch allows one to derive simultaneous constraints on the neutrino mass, MG and the DE equation of state. The new patch also makes it easy to implement alternative parameterizations of the MG functions and the DE equation of state.

\section{Discussion}
\label{sec:conclusion}

This paper presents a significant update of MGCAMB that features a consistent implementation of massive neutrinos and dynamical dark energy, as well as a new structure that renders the implementation of custom models easier. The new version also has no restriction on the value of the transition time at which the modifications to the linearized Einstein equations are switched on.

MGCAMB was the first publicly released modified Boltzmann solver for cosmological tests of gravity. Since its introduction in 2008 \cite{Zhao:2008bn}, MGCAMB has been used in over 100 works. A number of other codes have been introduced since, most notably ISiTGR \cite{Dossett:2011tn}, EFTCAMB \cite{Hu:2013twa,Raveri:2014cka} and hi\_class \cite{Zumalacarregui:2016pph}. Of them, ISiTGR is close to MGCAMB in its spirit, also introducing phenomenological modifications of equations of motion using two functions $Q$ and $R$ defined in our Appendix \ref{sec:QR}. EFTCAMB is based on the effective description of the background and perturbations solutions in general scalar-tensor theories \cite{Bloomfield:2012ff,Gleyzes:2013ooa,Bloomfield:2013efa}, while hi\_class uses an alternative effective description of perturbations in scalar-tensor theories on a fixed background \cite{Bellini:2014fua}. 

MGCAMB is best suited for model-independent constraints on $\mu$ and $\Sigma$, sometimes referred to as $G_{\rm matter}$ and $G_{\rm light}$, which are closely related to observables. The choice of the parameterizated forms of $\mu$ and $\Sigma$ can be informed by the QSA limit of particular types of modified gravity theories \cite{Pogosian:2016pwr}. One can also perform non-parametric reconstructions of $\mu$ and $\Sigma$ aided by a prior covariance derived from ensembles of modified gravity theories. Such priors can be obtained with the help of EFTCAMB as in Ref.~\cite{Espejo:2018hxa}.

With the present update, MGCAMB should remain a useful tool for cosmological tests of gravity, offering accuracy appropriate for data expected from the next generation surveys such as Euclid \cite{EUCLID} and LSST \cite{LSST}.

\acknowledgments

We thank Alireza Hojjati for co-developing and maintaining the earlier versions of MGCAMB. We also thank Marco Raveri, Matteo Martinelli and Meng-Xiang Lin for useful discussions. The work of AZ and LP is supported in part by the National Sciences and Engineering Research Council (NSERC) of Canada. AS acknowledges support from the NWO and the Dutch Ministry of Education, Culture and Science (OCW), and also from the D-ITP consortium, a program of the NWO that is funded by the OCW. GBZ is supported by NSFC Grants 1171001024 and 11673025, and the National Key Basic Research and Development Program of China (No. 2018YFA0404503).

\appendix
\section{Other parameterizations}
\label{sec:other_parameterizations}
\subsection{The $\mu, \Sigma$ parameterization}
As mentioned in Sec.~\ref{sec:framework}, rather than working with $\mu$ and $\gamma$, it can be beneficial to constrain $\mu$ and $\Sigma$, with the latter defined in Eq.~\eqref{eq:Sigma}. In this version of MGCAMB, we implement ($\mu,\Sigma$) by mapping it onto ($\mu,\gamma$) using $\gamma=2\Sigma-\mu$, which agrees with Eq.~\eqref{eq:Sigma} in the limit of negligible anisotropic stress ($\sigma \rightarrow 0$). In other words, we \emph{define} $\Sigma$ as
\be
\Sigma={1\over 2} \mu (1+\gamma).
\ee
The circumstances in which the difference between this definition and the one in Eq.~\eqref{eq:Sigma} can be important are not entirely clear to us. If necessary, it is relatively easy to add to MGCAMB a separate set of equations for the ($\mu,\Sigma$) parameterization based on Eq.~\eqref{eq:Sigma}.

\subsection{The $Q,R$ parameterization}
\label{sec:QR}
Another phenomenological parameterization was introduced in \citep{2010PhRvD..81h3534B} in which modifications of  gravity are encoded in functions $Q$ and $R$ defined through
\begin{gather}
k^2 \Phi = - 4 \pi G a^2 Q \rho \Delta, \\
k^2 \left( \Psi - R \Phi \right) = -12 \pi G Q a^2 (\rho + P) \sigma.
\end{gather}
This parameterization is consistently implemented in MGCAMB. The corresponding equation for $\dot{\eta}$ is
\begin{equation}
\begin{split}
\dot{\eta} &  = \frac{1}{2}\frac{1}{\frac{3}{2} Q a^2 \rho(1+w) + k^2} \biggl\{ Q k \rho q \left( 1+ \frac{3}{k^2} \left( \mathcal{H}^2 - \dot{\mathcal{H}} \right) \right) \\
& + \rho\Delta \left( \mathcal{H} Q(1-R) - \dot{Q}\right) \\ 
&+ k^2 \alpha \left( Q \rho(1+ w ) - 2 \frac{\mathcal{H}^2 - \dot{\mathcal{H}}}{a^2}\right) \biggr\},
\end{split}
\end{equation}
where the factor of $8 \pi G$ is absorbed into $\rho$ and $P$. For the ISW effect, we replace Eq.~\eqref{eqn:Psi_dot} with 
\begin{equation}
\dot{\Psi} = R \dot{\Phi} + \dot{R} \Phi - \frac{\dot{Q} \rho a^2 \Pi}{k^2} - \frac{Q (\rho a^2 \Pi)^{\cdot}}{k^2}.
\end{equation}

\section{Linear DES data}
\label{sec:linear_DES_data}

\begin{figure*}[tbh]
\includegraphics[width = .95\textwidth]{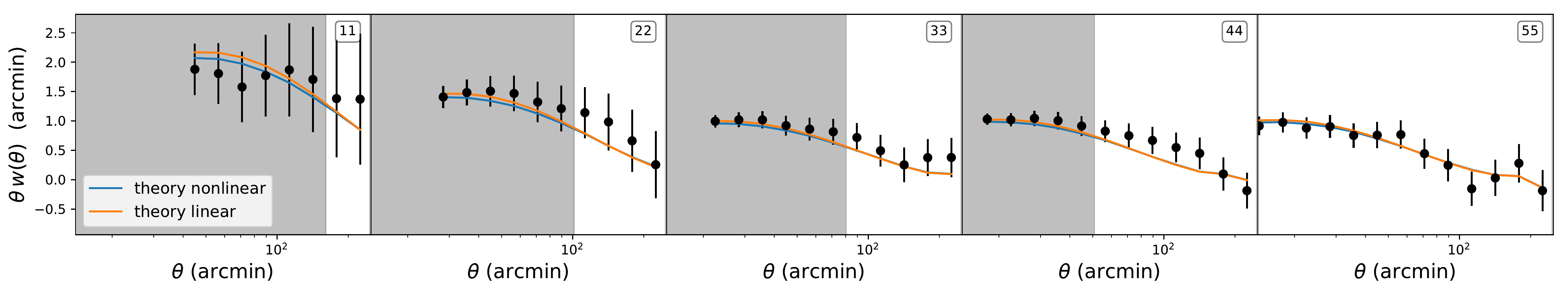}
\caption{\label{fig:DES_cuts_standard_2} An illustration of the ``standard'' cut on the DES dataset in the case of the galaxy-galaxy correlation function $w_{ij}(\theta)$ with $i=j$ in five redshift bins. The blue and orange lines represent the nonlinear and linear theoretical predictions, respectively. The data points inside the shaded regions are removed.}
\end{figure*}

Since the phenomenological parameterization implemented in MGCAMB has no prescription for nonlinear structure formation, in order to use the DES data, we remove the nonlinear data in the same way as was done in \citep{Ade:2015rim, Abbott:2018xao}. We define
\begin{equation}
\Delta \chi^2 \equiv ({\bf t}_{\rm NL} - {\bf t}_{\rm L})^T {\bf C}^{-1} ({\bf t}_{\rm NL} - {\bf t}_{\rm L}), 
\end{equation}
where ${\bf t}_{\rm NL}$ and ${\bf t}_{\rm L}$ represent the data vector containing the nonlinear and linear theory predictions, respectively, in the $\Lambda$CDM best-fit model. The nonlinear predictions are obtained using the Halofit model present in the default CAMB. We then find the data point that contributes the most to $\Delta \chi^2$ and remove it. We then repeat the procedure until $\Delta \chi^2$ is less than a threshold. We arbitrarily define three set of cuts on the data: a ``soft'' cut where  the $\Delta \chi^2$ threshold is 10, a ``standard'' cut with the threshold set to 5 and finally an ``aggressive'' cut with $\Delta \chi^2 < 1$. The number of data points removed are 88, 118 and 178, respectively. As an example, in Fig.~\ref{fig:DES_cuts_standard_2}  we show the ``standard'' cut applied to the DES galaxy-galaxy angular correlation function $w_{ij}(\theta)$, where $i,j$ label the redshift bins.  The blue and orange lines represent the nonlinear and linear theoretical predictions and the grey shaded lines show the data which is excluded by the above method.

\section{Galaxy Clustering - Weak lensing theory with Weyl potential.}
\label{sec:weyl_like}
Here we describe the modifications to the DES likelihood required for evaluating the weak lensing observables $\xi_+$, $\xi_-$ and $\gamma_t$. The standard DES likelihood in CosmoMC assumes $\Phi + \Psi = 2 \Phi$ and then relates the potential $\Phi$ to the density perturbation $\delta$ using the Poisson equation,
\begin{equation}
k^2\Phi = - \frac{3}{2 a} \left( \frac{H_0}{c} \right)^2 \Omega_m \delta.
\end{equation}
In MGCAMB, the Poisson equation is modified and $\Phi = \Psi$ does not hold. Hence, we compute the Weyl potential power spectrum directly,
\begin{equation}
P_{\rm Weyl}(k,z) = 2 \pi^2 k  \left( k^2 \frac{\tilde{\Phi}(z) + \tilde{\Psi}(z)}{2} \right)^2  P_{\cal R} (k),
\end{equation}
where the tilde quantities are the transfer functions at redshift $z$ and $P_{\cal R} (k)$ is the primordial power spectrum. The cosmic shear correlations $\xi_\pm$ are then given by
\begin{equation}
\xi_{\pm}^{ij}(\theta) = \frac{1}{2 \pi} \int d \ell \, \ell J_{0/4}(\theta \ell) P_{\kappa}^{ij}(\ell), 
\end{equation}
where $J_{0/4}(x)$  is the spherical Bessel function of order zero (fourth), $P_{\kappa}$ (in the Limber approximation) is given by
\begin{equation}
P_{\kappa}^{ij}(\ell) = \int_0^{\chi_H} d \chi \frac{q^i(\chi) q^j(\chi)}{\chi^2} P_{\rm Weyl} \left(\frac{\ell +1/2}{\chi}, \chi \right),
\end{equation}
where $q^i(\chi)$ is the lensing efficiency function,
\begin{equation}
q^i (\chi) = \chi \int_{\chi}^{\chi_H} d \chi^{\prime} \, n^i(\chi^{\prime}) \frac{\chi^{\prime} - \chi}{\chi^{\prime}},
\end{equation}
$n^i$ denotes the effective number density of galaxies normalized to one and the Weyl power spectrum is evaluated using the linear theory only. The nonlinear data is removed according to the procedure explained in App.~\ref{sec:linear_DES_data}.

Similarly, to calculate the tangential shear of background galaxies around foreground galaxies, we define the Weyl-matter power spectrum as
\begin{equation}
P_{\rm W/m} = 2 \pi^2 k \left( k^2 \frac{\tilde{\Phi}(z) + \tilde{\Psi}(z)}{2} \right) \tilde{\delta}_{\rm m}(k,z) P_{\cal R} (k).
\end{equation}
The tangential shear is then given by
\begin{equation}
\begin{split}
\gamma_t (\theta) =  b \int  \frac{d \ell}{2 \pi} \ell J_2(\theta \ell) \int dz \frac{g(z) n(z)}{\chi(z)} P_{\rm W/m}\left( \frac{\ell}{\chi}, \chi \right).
\end{split}
\end{equation}

\bibliography{paper.bib}

\end{document}